\documentclass[conference]{IEEEtran}
\IEEEoverridecommandlockouts
\usepackage{cite}
\usepackage{amsmath,amssymb,amsfonts}
\usepackage{algorithm}
\usepackage{algorithmic}
\usepackage{graphicx}
\usepackage{textcomp}
\usepackage{xcolor}
\usepackage{xspace}
\usepackage{soul}
\usepackage[utf8]{inputenc} % use UTF8 encoding
\usepackage{kotex} % use KoTeX package for Korean 

\def\BibTeX{{\rm B\kern-.05em{\sc i\kern-.025em b}\kern-.08em
    T\kern-.1667em\lower.7ex\hbox{E}\kern-.125emX}}
\begin{document}

\title{Quantum Neural Networks: Concepts, Applications, and Challenges}
% Quantum Deep Learning: A Road Ahead

\author{$^{\circ}$Yunseok Kwak, $^{\circ}$Won Joon Yun, $^{\circ}$Soyi Jung, and $^{\circ}$Joongheon Kim\\
\IEEEauthorblockA{$^{\circ}$Department of Electrical and Computer Engineering, Korea University, Seoul 02841, Republic of Korea 
\\
E-mails: 
\texttt{yskwak6937@gmail.com}, \texttt{ywjoon95@korea.ac.kr},\\ \texttt{jungsoyi@korea.ac.kr}, \texttt{joongheon@korea.ac.kr}
}
}
\maketitle

\begin{abstract}
Quantum deep learning is a research field for the use of quantum computing techniques for training deep neural networks. The research topics and directions of deep learning and quantum computing have been separated for long time, however by discovering that quantum circuits can act like artificial neural networks, quantum deep learning research is widely adopted. This paper explains the backgrounds and basic principles of quantum deep learning and also introduces major achievements. After that, this paper discusses the challenges of quantum deep learning research in multiple perspectives.
Lastly, this paper presents various future research directions and application fields of quantum deep learning. 
\end{abstract}

\section{Introduction}
As quantum computing and deep learning have recently begun to draw attentions, notable research achievements have been pouring over past decades. In the field of deep learning, the problems which were considered as their inherent limitations like gradient vanishing, local minimum, learning inefficiencies in large-scale parameter training are gradually being conquered~\cite{pieee202105park}. On the one hand, innovative new deep learning algorithms such as quantum neural network (QNN), convolutional neural network (CNN), and recurrent neural network (RNN) are completely changing the way various kinds of data are processed. Meanwhile, the field of quantum computing has also undergone rapid developments in recent years. Quantum computing, which has been recognized only for its potential for a long time, has opened up a new era of enormous potentials with the recent advances of variational quantum circuits (VQC). The surprising potentials of the variational quantum algorithms were made clear by solving various combinatorial optimization problems and the intrinsic energy problems of molecules, which were difficult to solve using conventional methods, and further extensions are considered to design machine learning algorithms using quantum computing. Among them, quantum deep learning fields are growing rapidly, inheriting the achievements of existing deep learning research. Accordingly, numerous notable achievements related to quantum deep learning have been published, and active follow-up studies are being conducted at this time. In this paper, we first briefly introduce the background knowledge, basic principles of quantum deep learning, and look at the current research directions. We then discuss the various directions and challenges of future research in quantum deep learning.

\subsection{Quantum Computing}
Quantum computers use qubits as the basic units of computation, which represent a superposition state between $|0\rangle$ and $|1\rangle$~\cite{app20choi,ictc19choi,icoin20choi,ictc20oh}. A single qubit state can be represented as a normalized two-dimensional complex vector, i.e.,
\begin{equation}
    \label{eq:qubit}    
|\psi\rangle = \alpha|0\rangle + \beta|1\rangle ,\|\alpha\|^2 + \|\beta\|^2 = 1
\end{equation}
and $\|\alpha\|^2$ and $\|\beta\|^2$ are the probabilities of observing $|0\rangle$ and $|1\rangle$ from the qubit, respectively. This can be also geometrically represented using polar coordinates $\theta$ and $\phi$, 
\begin{equation}\label{eq:bloch}    
|\psi\rangle = \cos(\theta/2)|0\rangle + e^{i\phi}\sin(\theta/2)|1\rangle, 
\end{equation}
where $0\leq\theta\leq\pi$ and $0\leq\phi\leq\pi$.
This representation maps a single qubit state into the surface of 3-dimensional unit sphere, which is called Bloch sphere. 
A multi qubit system can be represented as the tensor product of $n$ single qubits, which exists as a superposition of $2^n$ basis states from $|00...00\rangle$ to $|11...11\rangle$. Quantum entanglement appears as a correlation between different qubits in this system. For example, in a 2-qubit system $\frac{1}{\sqrt{2}}|00\rangle + \frac{1}{\sqrt{2}}|11\rangle$, the observation of the first qubit directly determines that of the second qubit. Those systems are controlled by quantum gates in a quantum circuit to perform a quantum computation on its purpose~\cite{icoin21choi,icoin21oh}.

Quantum gates are unitary operators mapping a qubit system into another one, and as classical computing, it is known that every quantum gate can be factorized into the combination of several basic operators like rotation operator gates and CX gate~\cite{electronics20choi}. 
Rotation operator gates $R_x(\theta), R_y(\theta), R_z(\theta)$ rotates a qubit state in Bloch sphere around corresponding axis by $\theta$ and CX gate entangles two qubits by flipping a qubit state if the other is $|1\rangle$. Those quantum gates utilizes quantum superposition and entanglement to take an advantage over classical computing, and it is well known that quantum algorithms can obtain an exponential computational gain over existing algorithms in certain tasks such as prime factorization~\cite{shor1999polynomial}. 
\section{Quantum Deep Learning}
\subsection{Variational Quantum Circuits (VQC)}
A variational quantum circuit (VQC) is a quantum circuit using rotation operator gates with free parameters to perform various numerical tasks, such as approximation, optimization, classification. An algorithm using a variational quantum circuit is called variational quantum algorithm (VQA), which is a classical-quantum hybrid algorithm because its parameter optimization is often performed by a classical computer. Since its universal function approximating property~\cite{biamonte2021universal}, many algorithms using VQC~\cite{cerezo2020variational} are designed to solve various numerical problems~\cite{farhi2014quantum,kandala2017hardware,app20choi,electronics20choi,apwcs21kim}. 
This flow led to many applications of VQA in machine learning and is also for replacing the artificial neural network of the existing model with VQC~\cite{schuld2019quantum,cong2019qcnn,bausch2020recurrent,dong2008quantum}. 
VQC is similar to artificial neural networks in that it approximates functions through parameter learning, but has differences due to the several characteristics of quantum computing. Since all quantum gate operations are reversible linear operations, quantum circuits use entanglement layers instead of activation functions to have multilayer structures. These VQCs are called quantum neural networks, and this paper will look at them through classification according to their structure and characteristics. 

\subsection{Quantum Neural Networks}
\begin{figure}[htp] \centering \includegraphics[width=1\columnwidth]{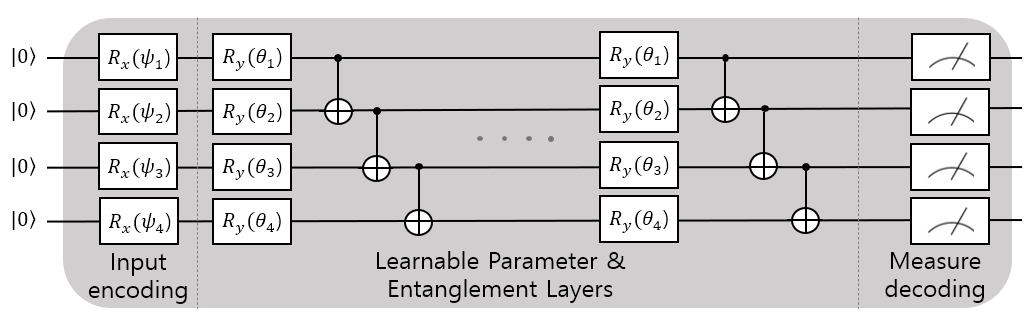} \caption{Illustration of QNN with the input $|\psi\rangle$, the parameter $\theta$ and linear entanglement structure.} \label{fig:QNN} \end{figure} 

In this section, we try to demonstrate how a basic quantum neural network(QNN) works with a simple example described in the Fig.~\ref{fig:QNN}. The way a QNN processes data is as follows. First, the input data is encoded into the corresponding qubit state of an appropriate number of qubits. Then, the qubit state is transformed through the parameterized rotation gates and entangling gates for a given number of layers. The transformed qubit state is then measured by obtaining expected value of a hamiltonian operator, such as Pauli gates. These measurements are decoded back into the form of appropriate output data. The parameters are then updated by an optimizer like Adam optimizer. A neural network constructed in the form of VQC can perform various roles in various forms, which will be explored as quantum neural networks. 

\subsubsection{Quantum Convolutional Neural Networks}
\begin{figure}[htp] \centering \includegraphics[width=1\columnwidth]{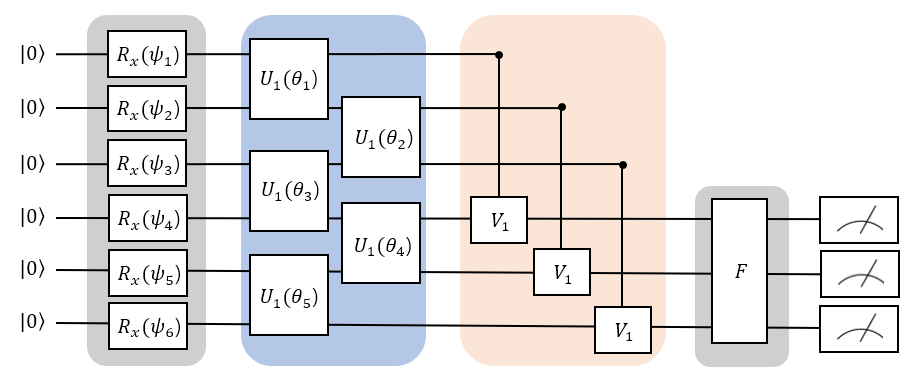} \caption{Illustration of QCNN with the input $|\psi\rangle$, the parameter $\theta$ with single convolution and pooling layer.} \label{fig:QCNN} \end{figure} 
Quantum convolutional neural network (QCNN) was proposed in \cite{cong2019qcnn}, implementing the convolution layer and pooling layer on the quantum circuits. According to the previous research results in \cite{ictc20oh,garg2020advances}, the QCNN circuit computation proceeds as follows. The first step is same as any other QNN models, encoding input data into a qubit state with rotation operator gates. Then the convolution layer with quasi-local unitary gates filters the input data into a feature map. The pooling layer with controlled rotation operators then downsizes the feature map. By repeating this process sufficiently, the fully connected layer acts on the qubit state as classical CNN models. Finally, the measurement of the qubit state is decoded into an output data with desired sizes. The circuit parameters are updated with gradient descent based optimizer after each measurements.

Unfortunaltely, in the current quantum computing environment~\cite{preskill2018quantum}, QCNN is difficult to perform better than the existing classical CNN. However, it is expected that the QCNN will be able to obtain sufficient computational gains over the classical ones in future quantum computing environment where larger-size quantum calculations are possible~\cite{cong2019qcnn,ictc20oh}.

\section{Future Work Directions and Challenges}
\subsection{Applications of Quantum Deep Learning to Reinforcement Learning}
There are many research results applying deep learning to reinforcement learning to derive optimal actions from a complex state space~\cite{mnih2013playing,twc201912choi,tvt201905shin,tvt202106jung}. However, reinforcement learning research using quantum deep learning~\cite{dong2008quantum,chen2020variational,jerbi2021variational} is still in its infancy. The current approach step is to replace the policy training network with a quantum neural network from the existing deep neural network, but there remains the possibility of many algorithms applying various ideas of classical deep reinforcement learning researches. In particular, if it is proved that quantum computational gains can be obtained through QNN in a situation of high computational complexity due to the complex Markov decision process environment, quantum reinforcement learning will open a new horizon for reinforcement learning research.

\subsection{Applications of Quantum Deep Learning to Communication Networks}
The QNN and quantum reinforcement learning algorithms can be used in various research fields, and this paper considers the applications in terms of communications and networks.
In terms of the acceleration of computation in fully distributed platforms, e.g., blockchain~\cite{isj202003saad,isj2021boo}, QNN can be used.

In addition, various advanced communication technologies such as Internet of Things (IoT)~\cite{jsac201811dao,isj2021dao}, millimeter-wave networks~\cite{tvt2021jung,jcn201410kim}, caching networks~\cite{tmc202106malik,twc201910choi,twc202012choi,twc202104choi}, and video streaming/scheduling~\cite{ton201608kim,jsac201806choi,tmc201907koo,tmc2021yi} are good applications of QNN and quantum reinforcement learning algorithms.

\subsection{Challenges}
\subsubsection{Gradient Vanishing}
Vanishing gradient is a crucial problem in quantum deep learning as of classical deep learning. The problem of gradient disappearance while backpropagating many hidden layers has been considered a chronic problem in deep neural network computation. Since quantum neural networks also use gradient descent method training their parameters as classical ones, they have to solve the same problem. Classical deep learning models solve this problem by utilizing an appropriate activation function, but quantum deep learning does not use an activation function, thus eventually, a different solution is needed. A former research~\cite{mcclean2018barren} called this quantum gradient vanishing pheonomena as barren plateaus, while proving that when the number of qubits increases, the probability of occurring barren plateaus increases exponentially. This can be avoided by setting good initial parameters in small-scale QNN, but it is unavoidable to deal with this problem when designing large-scale QNN. This is an open problem for which a solution is not yet clear.

\subsubsection{Near-Term Device Compatibility}
Noisy intermediate scale quantum (NISQ)~\cite{preskill2018quantum}, which means fewer qubits and a lot of computational error of near-term quantum devices, has already become a familiar term to quantum researchers. Many algorithms designed to implement quantum computational gains do not work at all in this NISQ environment, and are expected to be implemented at least several decades later. For example, a practical implementation of the Shor's algorithm requires at least thousand of qubits even without an error correction processes, current quantum devices have only a few tens of qubits with non-negligible computational error rate of several percent. However, due to the relatively small circuit depth and qubit requirements, VQA and QNN based on them are tolerant to these environmental constraints. Nevertheless, in order to increase the data processing capability of quantum neural network, it is necessary to consider near-term device compatibility. For example, using many multi-qubit controlling gates for quantum entanglement is theoretically thought to increase the performance of QNN, but it entails a large error rate and a complicated error correction process. Therefore, it is essential to design an algorithm regarding these tradeoffs in quantum deep learning research.

\subsubsection{The Quantum Advantage}
The term quantum supremacy may lead to the illusion that quantum algorithms are always better than classical algorithms performing the same function. However, given the inherent limitations of quantum computing, quantum computing benefits can only be realized through well-thought-out algorithms under certain circumstances. In fact, especially among variational quantum-based algorithms, only a few of them have proven their quantum advantage in a limited situation.

Due to the universal approximation property of QNN, it is known that quantum deep learning can perform most of the computations performed in classical deep learning~\cite{biamonte2021universal}. Nevertheless, if one approaches simply based on this fact without the consideration of quantum gain, the result may be much inefficient compared to the existing classical algorithm. Therefore, in designing a new QNN-based deep learning algorithm, it is necessary to justify it by articulating its advantages over the corresponding classical models.

\section{Conclusion}\label{sec:sec5}
This paper introduces the basic concepts of quantum neural networks and their applications scenarios in various fields. Furthermore, this paper presents research challenges and potential solutions of quantum neural network computation.

\section*{Acknowledgment}
This work was supported by the National Research Foundation of Korea (2019M3E4A1080391).
Joongheon Kim is a corresponding author of this paper.
\bibliographystyle{IEEEtran}
\bibliography{ref_quantum,ref_aimlab}
\end{document}